\documentclass{article}
\usepackage{graphicx}
\usepackage{authblk}

\begin{document}

	\title{Study of scaling laws in language families}
	\author[1,2]{M. R. F. Santos}
	\author[1]{M. A. F. Gomes}
	\affil[1]{Departamento de F\'isica, Universidade Federal de Pernambuco, 50670-901 Recife, PE, Brazil}
	\affil[2]{Instituto Federal de Educa\c{c}\~ao, Ci\^encia e Tecnologia de Pernambuco, 55560-000 Barreiros, PE, Brazil}
	\date{}

	\maketitle

\begin{abstract}
This article investigates scaling laws within language families using data from over six thousand languages and analyzing emergent patterns observed in Zipf-like classification graphs. Both macroscopic (based on number of languages by family) and microscopic (based on numbers of speakers by language on a family) aspects of these classifications are e\-xa\-mined. Particularly noteworthy is the discovery of a distinct division among the fourteen largest contemporary language families, excluding Afro-Asiatic and Nilo-Saharan languages. These families are found to be distributed across three language family quadruplets, each characterized by significantly different exponents in the Zipf graphs. This finding sheds light on the underlying structure and organization of major language fa\-mi\-lies, revealing intriguing insights into the nature of linguistic diversity and distribution.
\end{abstract}

\section{Introduction}
\label{section1}
\indent

Complex systems are extensively characterized by scaling symmetry and power law distributions. These phenomena are evident in a wide range of stu\-dies, such as those on cities \cite{Batty}, growth models \cite{Barabasi}, cellular automata \cite{Christensen}, and cognitive sciences \cite{Kello}, among many others. The language also exhibits characteristics of complex systems \cite{Balasubrahmanyan, Kretzschmar2015,  Stanisz} and currently several statistical laws related to linguistic studies are well established \cite{Altmann}.

Notedly interesting on the interface between statistical studies and linguistics, the Zipf's Law for written texts relates the word frequency with their corresponding rank and points to a robust power-law dependence between these variables with a scaling exponent close to unity \cite{Zipf49}. In recent decades, the expansion of available corpora and computational power has facilitated the study of Zipf's law across various language systems \cite{iCancho, Cancho, Jayaram, Baixeries, Moreno}.

Additionally, other types of  non-Zipfian scaling laws along several decades of variability were also found in the study of the distribution of living languages by Gomes et al. \cite{Gomes1999}, as well as by Santos and Gomes \cite{Santos2019}. Although these studies have explored the relations between linguistic diversity and certain geographical, demographic and economic factors, they have not delved into the investigation of scaling  within language families.

More than two decades ago, Zanette analyzed the distribution of language family sizes across seventeen families \cite{Zanette2001}. Wichmann later studied the distribution of family sizes using data from the fourteenth edition of Ethnologue \cite{Wichmann2005}, while Hammarstrom conducted a similar study based on his own language fa\-mi\-ly classification from the sixteenth edition of Ethnologue \cite{Hammarstrom2010}. In this paper, we first examine the distribution of language fa\-mi\-ly sizes using a more recent dataset. Differently from all previous studies, we then present an original ana\-lysis of the distribution of language sizes within each family, focusing on the fourteen largest language families.

The structure of this paper is as follows: in Section 2 we discuss the sca\-ling law emerging from the classification of language families according to the number of languages. In Section 3 we present the distribution of language sizes as measured by the number of speakers of 14 largest contemporary language families and discuss the appearing of quadruplets of language families. Section 4 ends with a brief summary of our conclusions.

\section{Macroscopic aspects}
\label{section2}
\indent

We study the data obtained from the digital twentieth edition of \textit{Ethnologue} \cite{Ethnologue}. This particular edition classifies 6711 living languages into 141 families. In our analysis presented in this section, similar to that carried out by Wichmann \cite{Wichmann2005}, were not included 388 languages classified in the special categories for constructed languages, creoles, sign languages, isolated languages, mixed languages, pidgins and unclassified languages. An important characteristic to note is the range in the number of languages in each family, which varies from one, as in the Caraj\'a family, to over a thousand languages, as in the Niger-Congo family. With regard to this difference in size between language families, Greenhill listed five possible explanations: family age, population size, technology (agriculture/language dispersal hypothesis), geography-and-ecology and social factors \cite{Greenhill}.

\begin{figure}[h]
\centering
\includegraphics[width=0.8\textwidth]{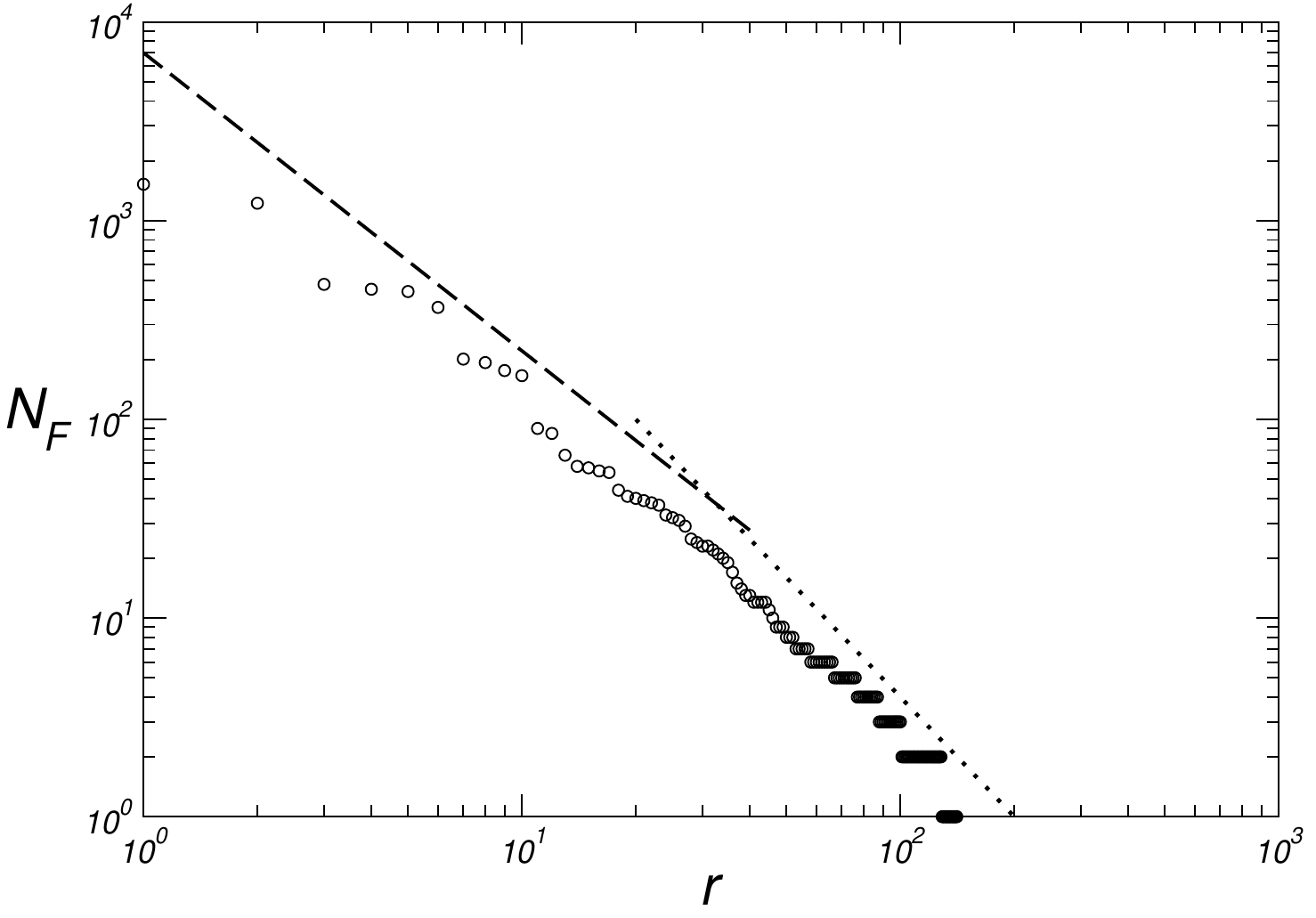}
\caption{Number of languages $N_F$ of each language family as a function of their rank $r$. The dotted (dashed) line with slope -2.0 (-1.5) corresponds to scaling behavior associated with stable distributions 
\cite{Takayasu}. The scaling exponents describe the data along approximately one decade of variability in the values of $r$.}
\label{figure_rank_families}
\end{figure}

We began by classifying the families in an orderly ranking similar to the classification of words carried out by Zipf in \textit{Human Behavior and the Principle of Least Effort} \cite{Zipf49}. Thus we assign the rank $r = 1$ to the Niger-Congo family, which is made up of 1526 languages, rank $r = 2$ to the Austronesian family, which is made up of 1224 languages, rank $r = 3$ to the Trans-New Guinea family, which is made up of 478 languages, and so on down the line. In this way we can write the cardinal size, i.e. the number of languages, $N_F$, that make up a family of classification $r$ according to
\begin{equation}
	N_F \sim r^{-\theta}.
\end{equation}
The resulting graph is shown in Figure \ref{figure_rank_families}. The choice of double logarithmic axes in this figure is justified so that the plotted points can be visualized linearly (a a useful introduction to visualization of that type of data was provided by Wichmann \cite{Wichmann2005} while Newman provides a more in-depth discussion of power laws \cite{Newman2005}).

In Figure \ref{figure_rank_families} can be viewed two different scaling behaviours associated with two distinct va\-lues for the exponent theta: a first region with $\theta = 1.5$ for intermediary values of rank and a second region with $\theta = 2.0$ for large values of rank. It is important to note that the dotted and dashed lines in Figure 1 (and subsequent figures) are not fits to the data. Instead, they represent our proposed assignments for the evolution of the data. The distribution of the size of language families that we report here is similar to the distribution curve of biological families, also called the Hollow Curve \cite{Greenhill}.

For large values of $r$, the exponent $\theta = 2.0$ refered in Figure \ref{figure_rank_families} is close to the value ($\theta = 1.905$) previously obtained from the fourteenth edition of \textit{Ethnologue} \cite{Wichmann2005}  but greater than the value ($\theta = 1.38$) obtained by Hammarstr\"om from their own language family classification from the sixteenth edition  of \textit{Ethnologue} \cite{Hammarstrom2010}. This last value, however, seems more akin to the the slope $1.5$ observed for the intermediary rank values in Figure \ref{figure_rank_families}. Our result contradicts Zanette \cite{Zanette2001} who analyzing a set of seventeen families from \textit{A Guide to the World's Languages} by Ruhlen \cite{Ruhlen}, proposed that the number of languages would decrease exponentially with the family rank.

Hammarstr\"om points out that since linguistic differentiation occurs mainly through human migration, the cardinal size of a family can be considered a measure of the diffusive spread of a family \cite{Hammarstrom2010} . From this perspective, we can understand that only a small number of language families are spread over large areas of the Earth's surface. At the same time, it is possible to understand that most families have a small geographical reach.

\section{Microscopic aspects}
\label{section3}
\indent

In Figure \ref{figure_rank_families} its possible see that only ten families have more than a hundred languages. It is worth asking whether the pattern observed in Figure \ref{figure_rank_families} is also observed within families. If so, this would imply that when classifying languages according to the linguistic population $N$ we should observe
\begin{equation}
	N \sim r^{-\kappa},
\end{equation}
where $r$ is the language classification and $\kappa$ is the characteristic exponent. Table \ref{table} shows the kappa value for the fourteen largest language families according to the number of speakers. In the following paragraphs we will discuss each of these families specifically and will show that, with the exception of the Afro-Asian and Nilo-Saharan families, the remaining twelve families are distributed in three quadruplets of language families grouped according those exponent of Zipf's distributions.

The largest language family, the Niger-Congo, is composed of 1526 languages and encompasses nearly all native languages in Africa below the Sahara and is characterized by $\kappa = 1.2$. Another family, consisting of over a thousand languages, is the Austronesian, with its 1224 languages scattered from Indonesia through the island of New Guinea to Easter Island. This family, originating from the region of Taiwan, has $\kappa = 1.6$. New Guinea is also home to the Trans-New Guinea family, composed of 478 languages and characterized by $\kappa = 1.1$. The Sino-Tibetan family has 452 languages but boasts over 380 times more speakers than the Trans-New Guinea family and it has $\kappa = 1.7$. The Indo-European family includes almost all languages in Europe and many languages in the Asian continent. Among the top twenty languages globally, eleven belong to this family, totalizing over three billion of speakers distributed across 440 languages with $\kappa = 1.7$.

\begin{table}
\centering 
\caption{Fourteen largest language families according to the number of languages $N_F$. The value $r$ indicates the classification of the language according to this ordering and $r_s$ indicates the classification of the family according to the linguistic population $N$. The value $\kappa$ is the exponent of the scaling law $N \sim r^{-\kappa}$.}
\begin{tabular}{cccccc}
\label{table}
  \textbf{$r$} & \textbf{Family} &  \textbf{$N_F$} &  \textbf{$N$ (in Millions)} &  \textbf{$r_s$} & \textbf{$\kappa$} \\ \hline
01 & Niger-Congo & 1475 & 458,90 & 03 & $1,2$ \\
02 & Austronesian & 1224 &  324,88 & 05 & $1,6$ \\   
03 & Trans-New Guinea & 478  & 3,55 & 21 & $1,1$ \\  
04 & Sino-Tibetan & 452  & 1355,71 & 02 & $1,7$ \\   
05 & Indo-European & 440  & 3077,11 & 01 & $1,7$ \\  
06 & Afro-Asiatic & 366  & 444,85 & 04 & $2,6$ \\   
07 & Nilo-Saharan & 201  & 50,33 & 12 & $1,4$ \\   
08 & Australian & 193  & 0,04 & 51 & $1,6$ \\ 
09 & Otomanguean & 176  & 1,68 & 24 & $1,1$ \\  
10 & Austro-Asiatic & 166  & 104,99 & 09 & $2,0$ \\   
11 & Tai-Kadai & 90 & 80,1 & 10 & $1,2$  \\   
12 & Dravidian & 85 & 228,1 & 06 & $2,0$  \\   
13 & Tupian & 66 & 6,2 & 19 & $2,1$  \\   
14 & Uto-Astecan & 58 & 1,9 & 22 & $2,1$  \\ 
\end{tabular} 
\end{table}

The 366 languages that make up the Afro-Asiatic family likely descend from the language spoken by human groups that migrated from the African continent to the Middle East over 50,000 years ago. It was in Phoenician, a language of this family, that the first phonetic alphabet was constructed. Unlike the five largest families characterized by two exponent values ($\kappa = 1.15 \pm 0.05$ and $\kappa = 1.65 \pm 0.05$), the Afro-Asiatic family has $\kappa = 2.6$, making this value the highest among the fourteen major language families. South of the Afro-Asiatic languages region lies another family whose $\kappa$ value is also distinct from those two characteristic of the five largest families: comprising over 200 languages, the Nilo-Saharan family has $\kappa = 1.4$.

The Australian family, although very diverse in the number of languages (193 in total), has experienced a significant decline in the number of speakers over the past centuries, making it the smallest family by this criterion among those discussed here. Australian languages with fewer than ten thousand speakers have $\kappa = 1.6$, which is the same value reported for the Austronesian family. The Otomanguean family, characteristic of the current Mexican territory, also experienced a population reduction after colonization processes and presents $\kappa = 1.1$. This value is identical to that reported for the Trans-New Guinea family.

The Austro-Asiatic language family, spoken from Southeast Asia to East India, ranks as the ninth-largest language family globally, with $\kappa = 2.0$. Southeast Asia is also home to the Tai-Kadai family, boasting about half the number of languages as the Austro-Asiatic family and that it's characterized by $\kappa = 1.2$. Moving to the Dravidian family, typical in South Asia, it boasts a linguistic po\-pu\-lation exceeding two hundred million speakers. Many of these languages persisted despite the expansion and dominance of Indo-European languages in the Indian region. The Dravidian family shares the same exponent as the Austro-Asiatic family, with $\kappa = 2.0$.

The processes resulting from colonisation irreversibly affected the distribution of languages on the American continent. Two of the language families most affected were the Tupian, in South America, and the Uto-Aztecan, in North America. These two families are characterised by $\kappa= 2.1$. Together with the Austro-Asiatic and Dravidian families, these two families constitute a quadruplet characterised by $\kappa = 2.05 \pm 0.05$.

\begin{figure}[h]
\centering
\includegraphics[width=0.8\textwidth]{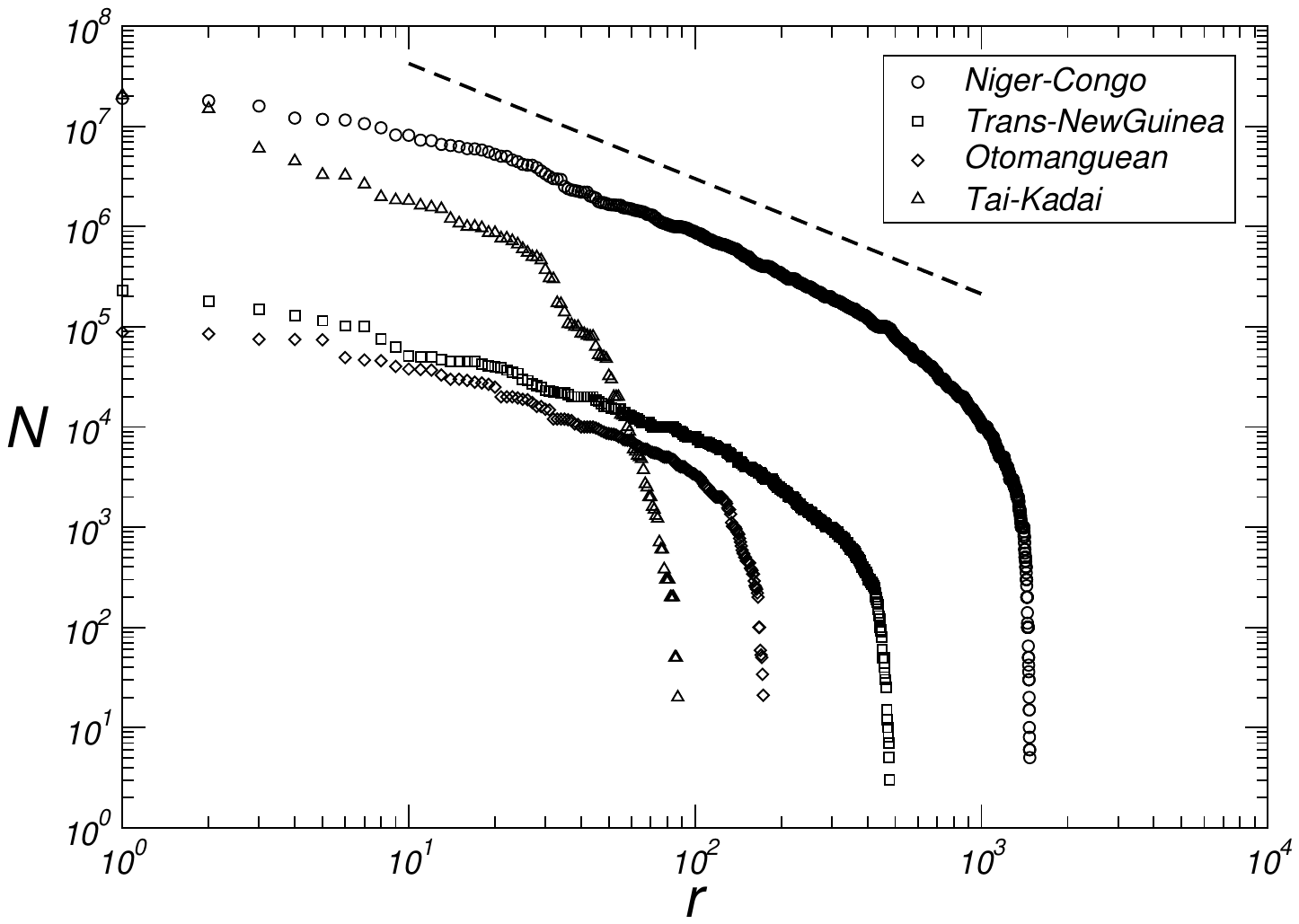}
\caption{Number of speakers $N$ by language as a function of rank $r$ for the Niger-Congo, Trans-New Guinea, Otomanguean and Tai-Kadai families. The dashed line provide guided to the eyes adjustments $N \sim r^{-\kappa}$ with $\kappa = 1.15$.}
\label{figure02}
\end{figure}

\begin{figure}
\centering
\includegraphics[width=0.8\textwidth]{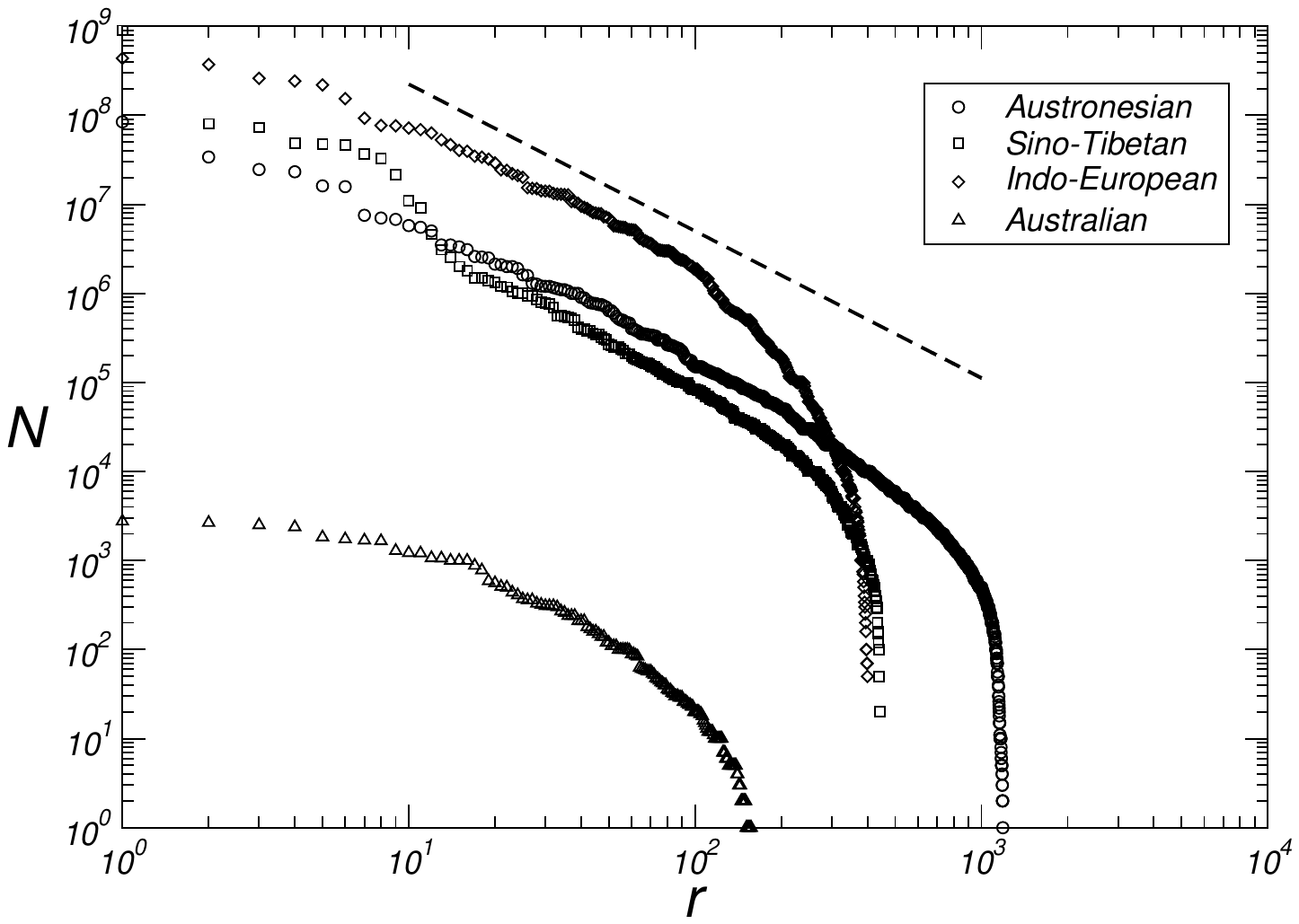}
\caption{Number of speakers $N$ by language as a function of rank $r$ for the Austronesian, Sino-Tibetan, Indo-European and Australian families. The dashed line provide guided to the eyes adjustments $N \sim r^{-\kappa}$ with $\kappa = 1.65$.}
\label{figure03}
\end{figure}

\begin{figure}
\centering
\includegraphics[width=0.8\textwidth]{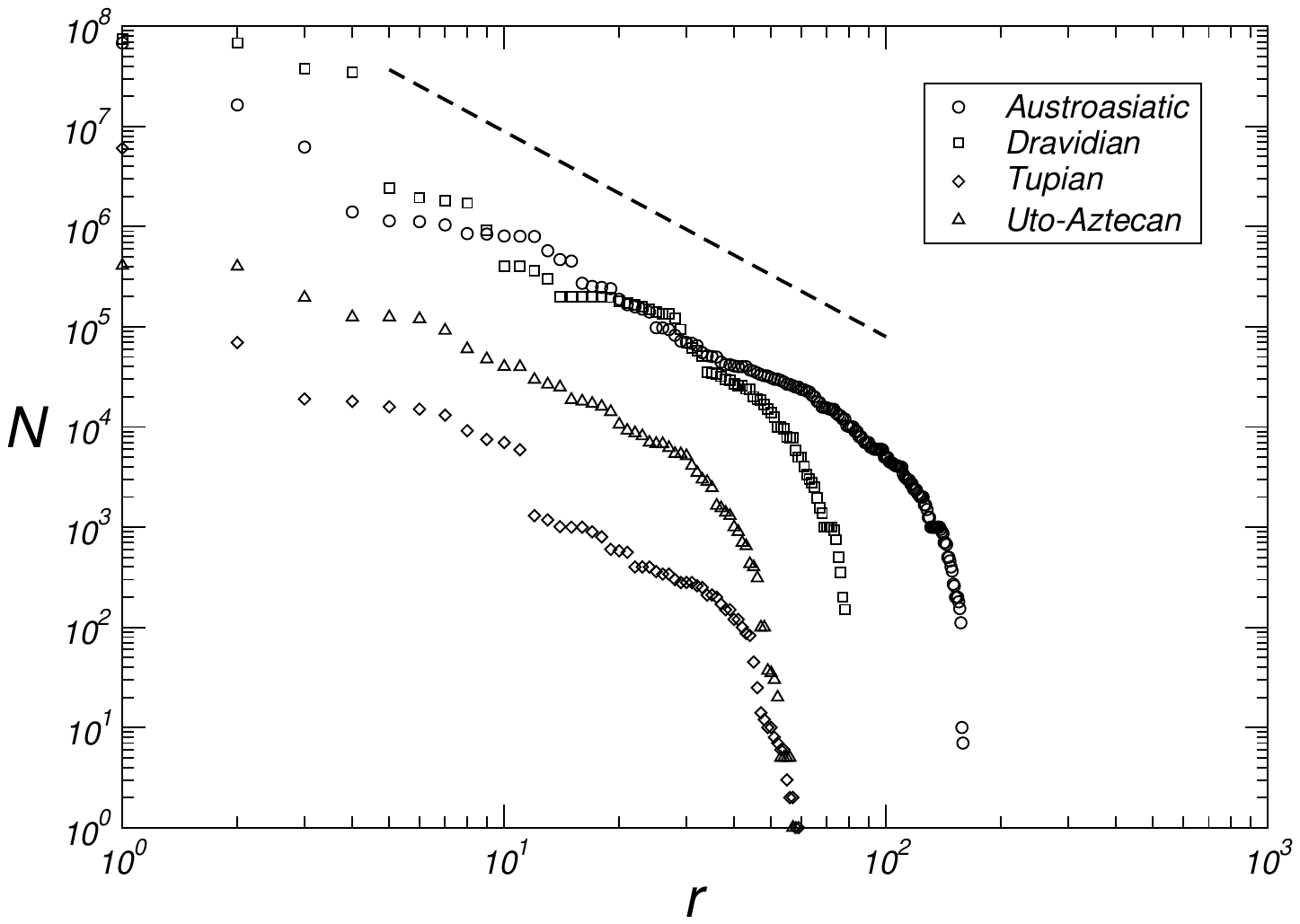}
\caption{Number of speakers $N$ by language as a function of rank $r$ for the Austroasiatic, Dravidian, Tupian and Uto-Aztecan families. The dashed line provide guided to the eyes adjustments $N \sim r^{-\kappa}$ with $\kappa = 2.05$.}
\label{figure04}
\end{figure}

Therefore, as discussed in the preceding paragraphs, with the exception of the Afro-Asiatic and Nilo-Saharan families, we have three quadruplets of language families grouped according to the exponent of the Zipf distributions. The first is composed of the Trans-New Guinea, Otomanguean, Niger-Congo and Tai-Kadai families with $\kappa = 1.15 \pm 0.05$ as seen in Figure \ref{figure02}. With $\kappa = 1.65 \pm 0.05$, the second quadruplet comprises the families Austronesian, Australian, Sino-Tibetan and Indo-European as seen in Figure \ref{figure03}. And finally the third quadruplet with the families Austro-Asiatic, Dravidian, Tupian and Uto-Aztecan has $\kappa = 2.05 \pm 0.05$ as seen in Figure \ref{figure04}.


\newpage
\section{Conclusions}
\label{section5}
\indent

Here we present power laws emerging from the distributions of both the size of linguistic families according the number of languages and the size of languages according their respective number of speakers within each of the fourteen largest families, based on methods for classifying and ordering data from more than six thousand languages. In contrast to previous studies, we show in Fi\-gu\-re \ref{figure_rank_families} that the cardinal size of a language family is related to the rank of the family by two exponents (1.5 and 2.0). The first for intermediate values of rank and the second for large values of rank. We then show that for twelve language families (Niger-Congo, Trans-New Guinea, Otomanguean, Tai-Kadai; Austronesian, Sino-Tibetan, Indo-European, Australian; Austroasiatic, Dravidian, Tupian, Uto-Aztecan) out of the fourteen largest families, we have three quadruplets of language families grouped according to the exponent of the Zipf distributions, namely, $\kappa = 1.15 \pm 0.05$ (Figure \ref{figure02}), $\kappa = 1.65 \pm 0.05$ (Figure \ref{figure03}) and $\kappa = 2.05 \pm 0.05$ (Figure \ref{figure04}). We believe that future research, with particular emphasis on detailed human migratory processes, should seek to understand (\textit{i}) why these twelve language families form statistically well-characterised quadruplets according to the values of the exponents, and (\textit{ii}) why the Afro-Asiatic and Nilo-Saharan families, both from the African continent, have different exponent values from those of the aforementioned quadruplets. Regarding this last aspect, it seems clear to us to expect that the continent where the greatest diversity of languages was originally generated should also present the grea\-test number of different emerging classes of statistical distributions that help to characterize this same diversity.

\section*{Acknowledgment}
\indent

M. A. F. Gomes acknowledges the financial support from the Brazilian Agency CAPES PROEX 23038.003069/2022-87, Grant \#0041/2022. \\

M. R. F. Santos acknowledges CAPES for doctoral scholarship at the beginning of this research, Instituto Federal de Pernambuco for license for the completion of the doctorate and Prof. Stanis\l{}aw Dro\.zd\.z for stimulating discussions at the CCS 2023.

\end{document}